\documentclass[prd,amsmath,amssymb,nofootinbib,superscriptaddress,twocolumn,10pt]{revtex4-1}

\pdfoutput=1

\usepackage{CJKutf8}
\usepackage{graphicx}
\usepackage{amsmath}
\usepackage{subfigure}
\usepackage{makecell}
\usepackage{siunitx}
\usepackage{booktabs}
\usepackage{float}

\usepackage[colorlinks=true, linkcolor=red, citecolor=blue]{hyperref}

\begin{document}

\title{Unifying topological, geometric, and complex classifications of black hole thermodynamics}

\author{Shi-Hao Zhang}
\email{pumazhang200@163.com}
\affiliation{Liaoning Key Laboratory of Cosmology and Astrophysics, College of Sciences, Northeastern University, Shenyang 110819, China}
\author{Shao-Wen Wei}
\email{weishw@lzu.edu.cn}
\affiliation{Lanzhou Center for Theoretical Physics, Key Laboratory of Theoretical Physics (Gansu) \& Key Laboratory of Quantum Theory and Applications (MOE), Lanzhou University, Lanzhou 730000, China}
\author{Jing-Fei Zhang}
\email{jfzhang@mail.neu.edu.cn}
\affiliation{Liaoning Key Laboratory of Cosmology and Astrophysics, College of Sciences, Northeastern University, Shenyang 110819, China}
\author{Xin Zhang}\thanks{Corresponding author}
\email{zhangxin@mail.neu.edu.cn}
\affiliation{Liaoning Key Laboratory of Cosmology and Astrophysics, College of Sciences, Northeastern University, Shenyang 110819, China}
\affiliation{MOE Key Laboratory of Data Analytics and Optimization for Smart Industry, Northeastern University, Shenyang 110819, China}
\affiliation{National Frontiers Science Center for Industrial Intelligence and Systems Optimization, 
	Northeastern University, Shenyang 110819, China}


\begin{abstract} 

Black hole thermodynamics has recently witnessed three distinct classification schemes: based on local geometric properties of the temperature function, global topological invariants, and Riemann surface foliations in the complex plane. We show that these schemes can be precisely mapped onto one another in the real domain via two dictionaries: one linking thermal stability to the monotonicity of the temperature curve, and the other connecting the number of black hole states to the foliation number of a Riemann surface. The number of extremal points of the temperature curve determines the classification in all three frameworks, tracing this unification to the critical point structure of the black hole solution space. As an illustration, several black holes demonstrate how counting extrema yields topological invariants and phase transition information. This unified framework simplifies black hole thermodynamic analysis and provides a foundation for exploring more complex black holes.
		
\end{abstract}
	

\maketitle

\section{Introduction}

General relativity is a gravitational theory based on Riemannian geometry. The study of the geometric properties of black holes, which are exact spacetime solutions predicted by general relativity, has long held a central position in gravitational theory research. In recent years, with the development of black hole thermodynamics, particularly the discovery of first-order phase transitions analogous to those of van der Waals fluids \cite{Kubiznak:2012}, attention has turned to whether these thermodynamic phenomena are fundamentally connected to the geometric properties of black holes.

Recent studies have revealed profound intrinsic connections between the thermodynamic, geometric, and topological properties of black holes. Investigations into the local geometric properties of the black hole temperature function have answered the question of how spacetime geometry encodes information about first-order phase transitions \cite{Zhang:2025inl}. Specifically, for black holes exhibiting a first-order phase transition, the solution space develops fold singularities, which cause branching behavior in the temperature function. This finding also explains the origin of the multivalued behavior with respect to temperature observed in previous studies for quantities such as the photon sphere radius \cite{PS01,PS02,PS03,PS04,PS05,PS06,PS07,PS08,PS09}, the Lyapunov exponent \cite{Guo:2022,Yang:2023,Lyu:2024,Kumara:2024,Du:2025,Shukla:2024,Gogoi:2024,Chen:2025xqc,R:2025gok,Awal:2025irl,Yang:2025,Kumar:2025kzt,Guo:2025pit,Bezboruah:2025udi,Ali:2025ooh,Xie:2025auj,MalikSultan:2026pok,Becar:2026epq,Cheng:2026dnd}, the Gaussian curvature \cite{Zhang:2025}, and other geometric quantities associated with extrinsic curvature \cite{Zhang:2025kqd}. Another significant advance is the interpretation of black holes as topological defects in the thermodynamic parameter space, offering a promising framework for investigating the universal properties of black hole thermodynamics through topology \cite{Wei:2022dzw,Wei:2024gfz}. Furthermore, studies that complex continue thermodynamic quantities into the complex domain have revealed new structures in the supercritical region, such as the Widom line \cite{Xu:2023vyj,Xu:2025jrk}.

Along these three seemingly independent directions, three representative classification schemes for black hole thermodynamics have gradually emerged. The first scheme is based on the locally defined geometric properties of the black hole temperature function \cite{Zhang:2025inl}. The second scheme relies on global topological invariants of black holes \cite{Wei:2024gfz}. The third scheme depends on the number of Riemann surface foliations in the complex plane, derived from the complex continued thermodynamic function \cite{Xu:2023vyj}.

These three schemes immediately raise several important questions. Do these three apparently independent classification schemes all reflect the same deep geometric property of black holes? If the answer is in the affirmative, then what change in the black hole solution space is responsible for simultaneously altering the topological charge, the number of critical points, and the foliation number of the Riemann surface? Furthermore, can we establish a correspondence among these three schemes that allows one to directly read off all the information of the other two frameworks starting from any one of them?

In this paper, we achieve a unification of these three schemes by establishing two core dictionaries, thereby addressed all the questions raised above. We demonstrate that the emergence of fold singularities within the black hole solution space is the fundamental cause of the change in topological properties and corresponds to the appearance of a foliated structure on the Riemann surface of the complex thermodynamic function. Through this unified perspective, one can select any single paradigm (for instance, the local geometric framework) and, by applying the corresponding dictionary, directly read off the other properties of the black hole, such as its topological number, black hole state systematic order, and the foliation number of its Riemann surface. Central to our unified framework is the recognition that the thermodynamic analysis of any black hole system should begin with its temperature function, which encodes both geometric and thermodynamic information. This approach will streamline future investigations into the thermodynamic behavior of more complex black holes and enhance our understanding of the geometric essence underlying black hole thermodynamics.

\section{Three frameworks}
Before embarking on the unification of the three frameworks, we briefly review them in this section, noting that they are distinct in style and independent in motivation. We first review the two frameworks formulated in the real domain, which are based on the global topological invariants of black holes and the local geometric properties, respectively.

The central idea of the former scheme is to treat black hole states as topological defects in the thermodynamic parameter space. By constructing a two-component vector field $\phi$, each black hole state corresponds to a zero point (defect) of $\phi$ \cite{Wei:2024gfz}. Each zero point can be assigned a winding number $w=+1,-1$, which corresponds to thermodynamic stability and instability, respectively. For degenerate zero points, the winding number can be defined as zero, but they are not counted as independent black hole states in the classification. For a detailed discussion of degenerate zero points in the context of topological defects, see Refs.~\cite{Wei:2022dzw,Wei:2024gfz}. The sum of all winding numbers is defined as the total topological number
\begin{align}
W = \sum_{i=1}^N w_i,
\end{align}
which is independent of external parameters (such as the cavity temperature) and is entirely determined by the intrinsic properties of the black hole system. Here $w_i$ represents the
winding number associated with the $i$-th zero point of $\phi$, with a total of $N$ zero points. Therefore, black holes can be classified into four categories, $W^{1-},~W^{0+},~W^{0-}$, and $~W^{1+}$, according to the value of $W$ and their asymptotic behavior. This scheme for the first time incorporates black hole thermodynamics into a topological classification framework, providing a global perspective for understanding the universal thermodynamic behavior of black holes.

The latter scheme originates from investigations into the mathematical origin of the synchronized multivaluedness exhibited by a series of physical or geometric quantities (such as the Lyapunov exponent \cite{Guo:2022,Yang:2023,Lyu:2024,Kumara:2024,Du:2025,Shukla:2024,Gogoi:2024,Chen:2025xqc,R:2025gok,Awal:2025irl,Yang:2025,Kumar:2025kzt,Guo:2025pit,Bezboruah:2025udi,Ali:2025ooh,Xie:2025auj,MalikSultan:2026pok,Becar:2026epq,Cheng:2026dnd}, photon sphere radius \cite{PS01,PS02,PS03,PS04,PS05,PS06,PS07,PS08,PS09}, and curvature \cite{Zhang:2025,Zhang:2025kqd}) when a black hole undergoes a first-order phase transition. Ref.~\cite{Zhang:2025inl} demonstrates that the mathematical origin of this multivaluedness lies in the temperature function $T(r_h)$ possessing two non-degenerate critical points. Based on this observation, they propose a classification of black holes according to the number of extremal points on the $T(r_h)$ curve: black holes with two extremal points are classified as class $A2$ (exhibiting a first-order phase transition), those with one extremal point as class $A1$, and those with no extremal points as class $B$. This scheme reduces the diagnosis of complex phase transition phenomena to the intuitive operation of counting the number of extremal points on the temperature curve, thereby revealing the local geometric origin of phase transition phenomena. In this paper, we refine class $A1$ into the $A1^+$ class, in which the temperature function has a single maximum, and the $A1^-$ class, in which it has a single minimum. Similarly, we refine class $B$, which has no extremum, into the $B^+$ class, where the temperature curve is monotonically increasing, and the $B^-$ class, where it is monotonically decreasing, in order to better align with the other schemes.

The third scheme emerges from explorations of black hole thermodynamics in the complex domain. By analytic continuation, this scheme extends the real generalized free energy to the complex plane, yielding a complex function $\psi(z)$ (obtained by differentiating the generalized free energy, where $z$ is the complexified horizon radius). Black hole states correspond to the zero points of $\psi(z)$ on the positive real axis. Using the Argument Principle from complex analysis, the winding number along a closed contour enclosing all physical zero points can be computed; if the contour contains no poles, this winding number is precisely the local maximum winding number, which directly corresponds to the number of foliations of the Riemann surface associated with $\psi(z)$. Ref.~\cite{Xu:2023vyj} reveals that van der Waals type phase transitions are closely related to the number of Riemann surface foliations (for example, a three foliations Riemann surface corresponds to a first-order phase transition). This framework not only successfully extends black hole thermodynamics to the complex domain but also paves the way for exploring new structures such as the Widom line in the supercritical region by analyzing the distribution of zero points of the partition function in the complex plane \cite{Xu:2025jrk}.

As will be shown in the following two subsections, although the three frameworks described above differ in both motivation and language, once the corresponding dictionaries are established, these three frameworks can be unified.

\section{Dictionary and correspondence table}
In this section, we use two dictionaries to connect the three frameworks and present a correspondence table.
\subsection{Dictionary between local and global}

In black hole thermodynamics, the heat capacity is defined as
\begin{align}
\mathcal{C} = T \frac{\partial S}{\partial T} = T \frac{\partial S}{\partial r_h} \left( \frac{\partial T}{\partial r_h} \right)^{-1},
\end{align}
where $r_h$ is the horizon radius and $S$ is the entropy of the black hole. Since $T > 0$ and $\frac{\partial S}{\partial r_h} > 0$ for a general case, the sign of the heat capacity $\mathcal{C}$ is determined solely by the sign of $\frac{\partial T}{\partial r_h}$, which leads
\begin{align}
\operatorname{sign}(\mathcal{C}) = \operatorname{sign}\left(\frac{\partial T}{\partial r_h}\right).\label{c}
\end{align}
For a local stable black hole state, the heat capacity is positive; otherwise, it is negative.

Now we demonstrate how Eq.~(\ref{c}) serves as a bridge connecting the local and global schemes. We will proceed by analyzing in two directions: from the local scheme to the global scheme, and from the global scheme to the local scheme. Let the minimum horizon radius be denoted by $r_m$ (with horizon radius $r_h$), and let the temperature $T(r_h)$ be a real analytic function on the interval $(r_m,~\infty)$ with $r_m > 0$. All critical points of $T(r_h)$ are non-degenerate (for a single variable function, the requirement of a non-vanishing second derivative for a first-order phase transition ensures non-degeneracy and the presence of critical points). The entropy $S$ is strictly increasing on $(r_m,~\infty)$.

We begin from the local scheme, taking the $A2$ class black hole as an example. Its $T(r_h)$ curve possesses two non-degenerate critical points $r_1$ and $r_2$, which are, respectively, a local maximum $T(r_1)$ and a local minimum $T(r_2)$. On the monotonic interval $(r_m,~r_1)$, we have $T' > 0$; on $(r_1,~r_2)$, $T' < 0$; and on $(r_2,~\infty)$, $T' > 0$, where $T'=\frac{\partial T}{\partial r_h}$. This implies that the heat capacities on the three branches are, respectively, $\mathcal{C} > 0$, $\mathcal{C} < 0$, and $\mathcal{C} > 0$, with corresponding winding numbers $w_s=+1$ for the small black hole, $w_m=-1$ for the intermediate black hole, and $w_l=+1$ for the large black hole. The total topological number is therefore $W = (+1)+(-1)+(+1)=+1$. Thus, the $A2$ class in the local scheme corresponds to the $W^{1+}$ class in the global scheme, with the winding number order $\left[+,~-,~+ \right]$.

It is worth noting that for the $B^+$ class in the local scheme, which has no extremal points, we have $T' > 0$ on the entire interval $(r_m,~\infty)$, so $T(r_h)$ is monotonically increasing. Consequently, it possesses only one stable black hole state, with both the innermost and outermost states being stable. This corresponds to the $W^{1+}$ class in the global scheme, with the winding number order $\left[+\right]$. This shows that the $W^{1+}$ class in the global scheme actually encompasses two distinct types of black hole in the local scheme: the $A2$ class (with three black hole states) and the $B^+$ class (with only one black hole state).

Similarly, it can be demonstrated that the $A1^{+},~A1^{-}$ and $B^-$ class black holes in the local scheme, respectively, correspond to the $W^{0+},~W^{0-}$ and $W^{1-}$ class black holes in the global scheme.

If we instead start from the global scheme, taking the $W^{1+}$ class as an example, its innermost and outermost states are both stable, with the winding number orders $\left[+,~(-,~+),~\dots,~(-,~+)\right]$. From the perspective of stability, if only one additional pair of black hole states appears, the winding number order becomes $\left[+,~-,~+ \right]$, corresponding to local stable, unstable, and stable black hole states in the sequence. Based on the preceding analysis of the sign of the heat capacity, this corresponds to three branches with $T' > 0$, $T' < 0$, and $T' > 0$, respectively. Such a geometric structure implies that the $T(r_h)$ curve must possess one local maximum and one local minimum. This corresponds to the $A2$ class in the local scheme, and such black holes can undergo a first-order phase transition. If no additional paired black hole states exist, the winding number order is $\left[+\right]$, indicating no branching behavior. In this case, $T(r_h)$ is a monotonically increasing function on its domain with $T' > 0$, which corresponds to the $B^+$ class.

Similarly, it can be verified that the $W^{0+}$, $W^{0-}$, and $W^{1-}$ class black holes in the global scheme correspond to the $A1^{+}$, $A1^{-}$, and $B^{-}$ class black holes in the local scheme, respectively.

\subsection{Dictionary between real and complex domains}

In this subsection, we formulate the real analysis scheme and the complex analysis scheme in terms of the following two core statements. If these statements are equivalent, then the real analysis scheme and the complex analysis scheme are equivalent in the real domain. The extended real analysis framework posits that the temperature function $T(r_h)$ of a black hole possesses $n-1$ extremal points (non-degenerate critical points), thereby giving rise to $n$ branches in the parameter space. The complex analysis framework, by contrast, takes the view that, according to the Argument Principle, the local maximum winding number of the complex analytic function $\psi(z)$ is $n$, which corresponds to a Riemann surface with $n$ foliations.

Following the conclusion of the real analysis framework, if $T(r_h)$ has $n-1$ extremal points, then there exists a temperature interval $(T_{min},~T_{max})$ such that the equation $T(r_h) = T_0$ has $n$ positive real roots $(r_1,~r_2,~\dots, r_n)$; that is, there are $n$ black hole states, where $n$ is a positive integer. Since $T(r_h)$ is analytic for $r_h > 0$, following the method in Ref.~\cite{Xu:2023vyj}, we can extend it to a complex function $T(z)$ and further construct the complex analytic function
\begin{align}
\psi(z) = \frac{dU}{dS},
\end{align}
where $U=U(z)$ is the complexified free energy and $S=S(z)$ is the complexified entropy, $z=x+iy$, $x$ and $y$ are the real and imaginary parts of the complex argument $z$, respectively. Choose a sufficiently large simple closed contour $C$ in the complex plane that encloses these $n$ real roots and contains no poles. Since there are no poles inside $C$, by the Argument Principle we have
\begin{align}
w = \frac{1}{2\pi i} \oint_C \frac{\psi'(z)}{\psi(z)} \, dz = n,
\end{align}
so the local maximum winding number is $n$, corresponding to a $n$ foliations Riemann surface. Here, $\psi'(z)$ denotes the derivative with respect to $z$.

Conversely, within the complex analysis framework, if the local maximum winding number of $\psi(z)$ is $n$, then in the real domain there exists some $T_0$ such that the equation $T(r_h)=T_0$ has $n$ distinct positive real roots. On any interval formed by two adjacent roots $r_{i-1}$ and $r_{i}$ (with $i \leq n$), Rolle's theorem guarantees the existence of at least one point $\xi_i \in [r_{i-1},~r_{i}]$ such that $T'(\xi_i) = 0$. Since the local maximum winding number $n$ represents the maximum number of real roots attainable by the equation, $\xi_i$ must be the unique extremal point on that interval; otherwise, a finely tuned adjustment of $T_0$ would yield additional real roots, leading to a contradiction. Consequently, $n$ roots imply $n-1$ extremal points, which is precisely the multivaluedness discussed in the real analysis scheme \cite{Zhang:2025inl}.

Thus, there are two core bridges connecting the real and complex analyses. The first is the complex continuation of $T(r_h)$ to the complex domain as $T(z)$, from which the complex analytic function $\psi(z)$ is constructed. The second is the recognition that the black hole states $r_h$ on the real axis correspond precisely to the zeros of $\psi(z)$ on the real axis. It is this correspondence that directly links the number of extremal points in the real domain to the number of Riemann surface foliations in the complex domain via the Argument Principle.

\subsection{Correspondence table}

The correspondence among the different schemes is summarized in Table~\ref{classification}.

\begin{table*}[htbp]
\centering
\setlength{\tabcolsep}{6pt}
\begin{tabular*}{\textwidth}{c@{\extracolsep{\fill}}cccccc}
\hline\hline
\multicolumn{1}{c}{} & \multicolumn{2}{c}{Phase transition} & \multicolumn{3}{c}{Topology} & \multicolumn{1}{c}{Complex analysis} \\
\cmidrule(lr){2-3} \cmidrule(lr){4-6} \cmidrule(lr){7-7}
Local class & Extrema & 1st order & $W$ & Global class & Order & Riemann surface \\
\midrule
$A2$    & 2 & Yes & $+1$ & $W^{1+}$ & $[+, -, +]$ & 3 foliations \\
$A1^{+}$ & 1 & No  & $0$  & $W^{0+}$ & $[+, -]$    & 2 foliations \\
$A1^{-}$ & 1 & No  & $0$  & $W^{0-}$ & $[-, +]$    & 2 foliations \\
$B^{+}$  & 0 & No  & $+1$ & $W^{1+}$ & $[+]$       & 1 foliation  \\
$B^{-}$  & 0 & No  & $-1$ & $W^{1-}$ & $[-]$       & 1 foliation  \\
\bottomrule
\end{tabular*}
\caption{Correspondence among the local geometric classification, the global topological classification, and the complex analytic classification. For each black hole class, the number of extremal points of the temperature curve, the presence of a first-order phase transition, the topological number $W$, the winding number order, and the number of Riemann surface foliations are listed.}
\label{classification}
\end{table*}

Thus, by simply plotting the temperature curve $T(r_h)$ for any black hole and counting the number of extremal points, one can immediately read off its topological classification, winding number order, number of Riemann surface foliations, and whether a first-order phase transition occurs using the dictionaries established above. This significantly simplifies the analysis of black hole thermodynamics and provides a unified theoretical tool for subsequent investigations of more complex black hole systems.

It is noteworthy that a more precise classification of black holes at the thermodynamic critical point is necessary, where the non-degenerate branch structure breaks down and critical exponents are governed by the scaling of thermodynamic potentials rather than by the temperature function alone. However, this lies beyond the scope of the present work and will be addressed in future studies.

\section{Classification results for several black holes}
In this section, we examine the monotonic behavior of the temperature function for several black holes in the canonical ensemble, classify them, and present a summary table.
\subsection{Reissner-Nordstr\"om--AdS black hole}

We first take the Reissner-Nordstr\"om--AdS (RN--AdS) black hole as an example to illustrate what properties can be read off from the dictionaries established above.

For the RN--AdS black hole, the metric function is given by
\begin{align}
f_R(r)= 1 - \frac{2M}{r} + \frac{Q^2}{r^2} + \frac{r^2}{\ell^2},
\end{align}
where $Q$ denotes the black hole's charge, $M$ is the ADM mass of the black hole and $\ell$ is the AdS radius.
Setting $T'(r_h)=0$ yields
\begin{align}
-r_h^2 + 3Q^2 + \frac{3r_h^4}{\ell^2} = 0.\label{de}
\end{align}
From the discriminant of this equation, for the equation to have two distinct positive real roots (i.e., the $T(r_h)$ curve exhibits two extremal points and three black hole branches), we have $\ell>6Q$. In this case, the black hole undergoes a first-order phase transition and thus belongs to the $A2$ class. The critical charge is given by $Q = \ell/6$.

Using the dictionary between the local and global schemes, the RN--AdS black hole ($A2$ class) corresponds to the $W^{1+}$ class black hole, with topological number $W=1$ and winding number order $\left[+,~-,~+ \right]$. By complex continuing the function $T(r_h)$ to the complex domain, the corresponding meromorphic function $\psi(z)$ exhibits a Riemann surface with three foliations in the complex plane. Note that when $\ell$ is very small, the third term in Eq.~(\ref{de}) becomes dominant, and Eq.~(\ref{de}) admits only the trivial solution $r_h=0$, without two distinct positive real roots. In this case, the temperature function approximates $T\approx\frac{3r_h}{4\pi\ell^2}$, which is strictly monotonically increasing in $r_h$. Therefore, under such parameters, the RN--AdS black hole belongs to the $B^+$ class in the local scheme, corresponding to the $W^{1+}$ class in the global scheme with the winding number order $\left[+ \right]$, and the corresponding Riemann surface of $\psi(z)$ in the complex plane is one foliation. This indicates that, from a global perspective, although $W$ remains unchanged, the local behavior of the system changes, ultimately leading to a significantly different thermodynamic behavior (no first-order phase transition). For more black hole models, see next subsection. 

\subsection{Other black holes}

\subsubsection{Hayward--AdS}

As a regular black hole that avoids the singularity, the Hayward--AdS black hole has the metric function
\begin{align}
f = 1 - \frac{2Mr^2}{g^3 + r^3} + \frac{r^2}{\ell^2},
\end{align}
where $g$ is the magnetic charge, $M$ is the ADM mass of the black hole, and $\ell$ is the AdS radius. The temperature is given by
\begin{align}
T = \frac{(r_h^3 - 2g^3) + \frac{3r_h^5}{\ell^2}}{4\pi r_h(g^3 + r_h^3)}.
\end{align}
To facilitate the analysis of the relationship between the behavior of the temperature function and the extensive quantities, we introduce the following dimensionless rescaling (which does not change the essential behavior of the function $T(r_h)$):
\begin{align}
\tilde{r}_h = \frac{r_h}{\ell}, \quad \tilde{g} = \frac{g}{\ell}, \quad \tilde{M} = \frac{M}{\ell},  \quad \tilde{T} = T\ell.
\end{align}
At the critical point, $\tilde{g}_c=0.1423$, $\tilde{r}_{hc}=0.4356$.

\begin{figure}
\centering
\includegraphics[width=0.4485\textwidth, height=0.22425\textheight]{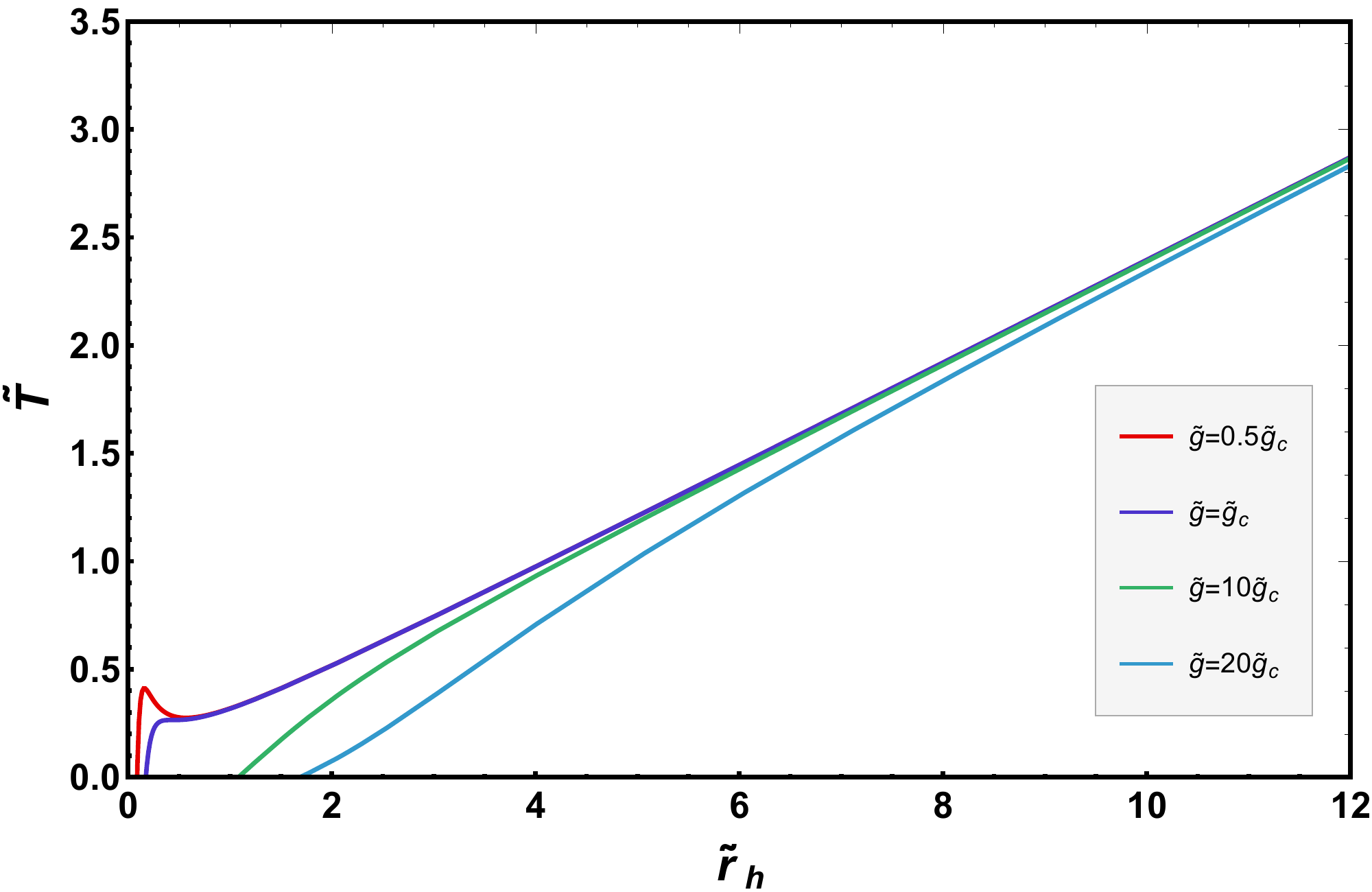} \hspace{1cm}
\caption{\label{1} The temperature function $\tilde{T}(\tilde{r}_h)$ of the Hayward--AdS black hole shown in the $\tilde{T}-\tilde{r}_h$ plane for different values of $\tilde{g}$.}
\end{figure}

When the Hayward--AdS black hole satisfies the critical condition (small $\tilde{g}$), its temperature curve exhibits two extremal points, and therefore it belongs to the $A2$ class. Note that when $\tilde{g}$ is large,  $\tilde{T}(\tilde{r_h})$ becomes a monotonically increasing function, rendering the Hayward--AdS black hole a $B^+$ class black hole. The behavior of the temperature function for different values of $\tilde{g}$ is shown in Fig.~\ref{1}.

\subsubsection{Hayward}
\begin{figure}
\centering
\includegraphics[width=0.4485\textwidth, height=0.22425\textheight]{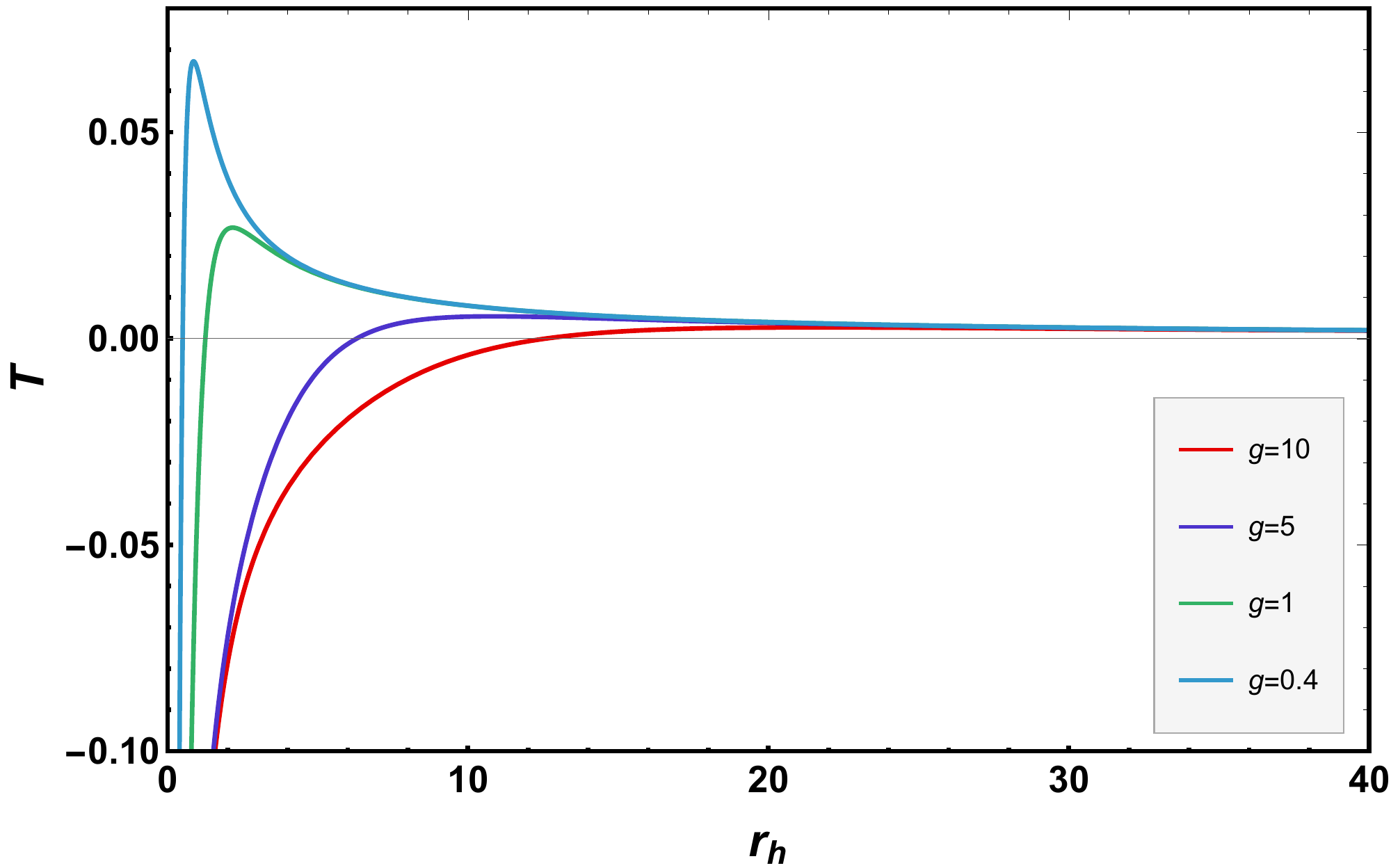} \hspace{1cm}
\caption{\label{2} The temperature function $T(r_h)$ of the Hayward black hole shown in the $T-r_h$ plane for different values of $g$.}
\end{figure}

The Hayward black hole is a regular black hole without a cosmological constant, with the metric function
\begin{align}
f = 1 - \frac{2Mr^2}{g^3 + r^3},
\end{align}
and the temperature
\begin{align}
T = \frac{(r_h^3 - 2g^3)}{4\pi r_h(g^3 + r_h^3)}.
\end{align}
As shown in Fig.~\ref{2}, regardless of the value of the control parameter $g$ (the magnetic charge, assumed to be positive), the temperature function of the Hayward black hole always possesses a single maximum, and thus belongs to the $A1^+$ class.

\subsubsection{Schwarzschild--AdS}
\begin{figure}
\centering
\includegraphics[width=0.4485\textwidth, height=0.22425\textheight]{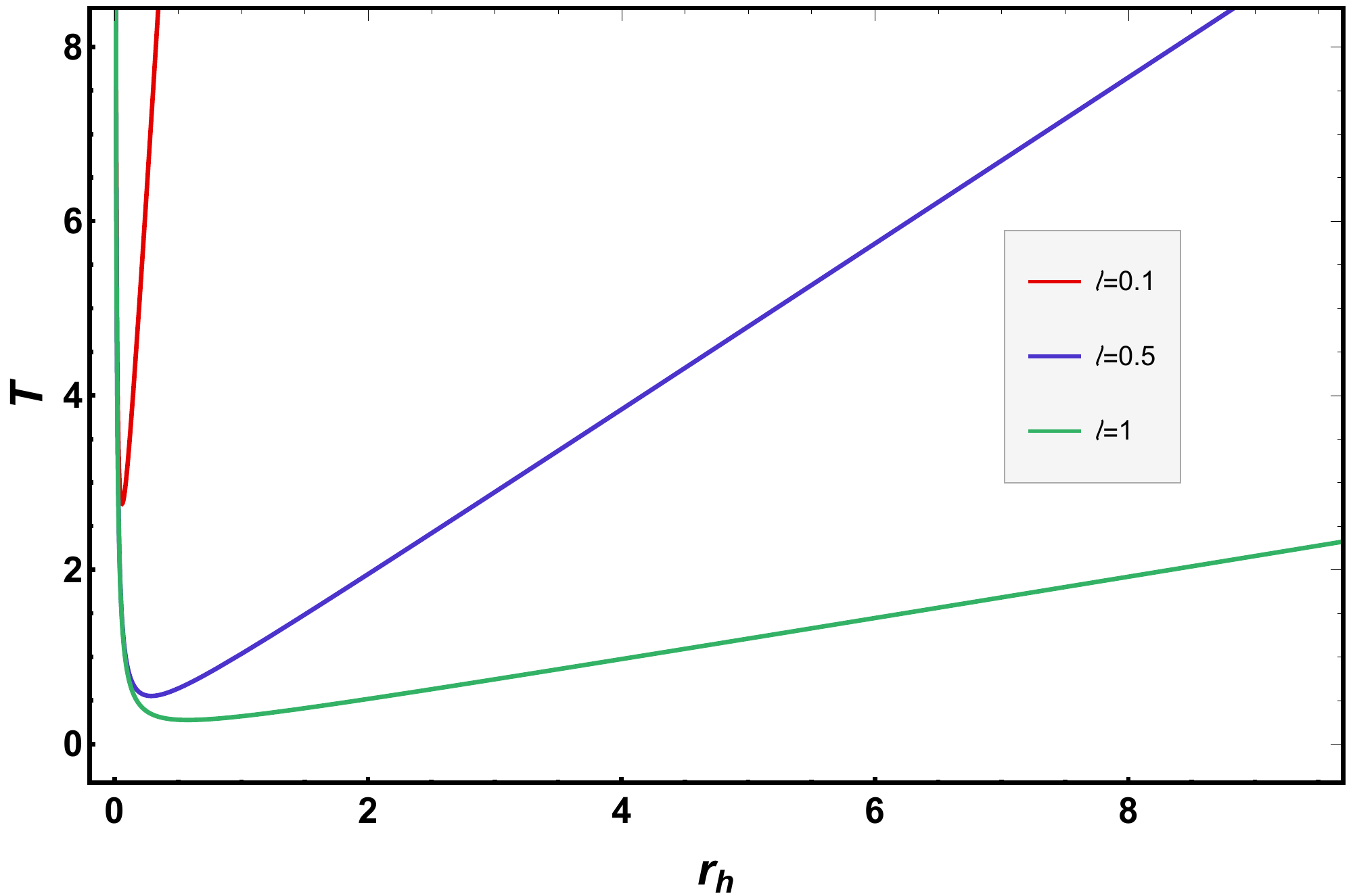} \hspace{1cm}
\caption{\label{3} The temperature function $T(r_h)$ of the Schwarzschild--AdS black hole shown in the $T-r_h$ plane for different values of $\ell$.}
\end{figure}

Schwarzschild--AdS is an uncharged static spherically symmetric black hole, with the metric function
\begin{align}
f = 1 - \frac{2M}{r} + \frac{r^2}{\ell^2},
\end{align}
and the temperature
\begin{align}
T = \frac{1}{4\pi} \left( \frac{1}{r_h} + \frac{3r_h}{\ell^2} \right).
\end{align}
As shown in Fig.~\ref{3}, when $\ell$ is sufficiently small, the term $\frac{3r_h}{\ell^2}$ dominates, so the temperature is a monotonically increasing function, and the Schwarzschild--AdS black hole belongs to the $B^+$ class. It is worth noting that, because $1/r_h$ does not vanish, $T(r_h)$ is not strictly increasing over its entire domain. Moreover, the temperature function can also exhibit a single minimum, in which case the Schwarzschild--AdS black hole falls into the $A1^-$ class.

\subsubsection{Schwarzschild}
\begin{figure}
\centering
\includegraphics[width=0.4485\textwidth, height=0.22425\textheight]{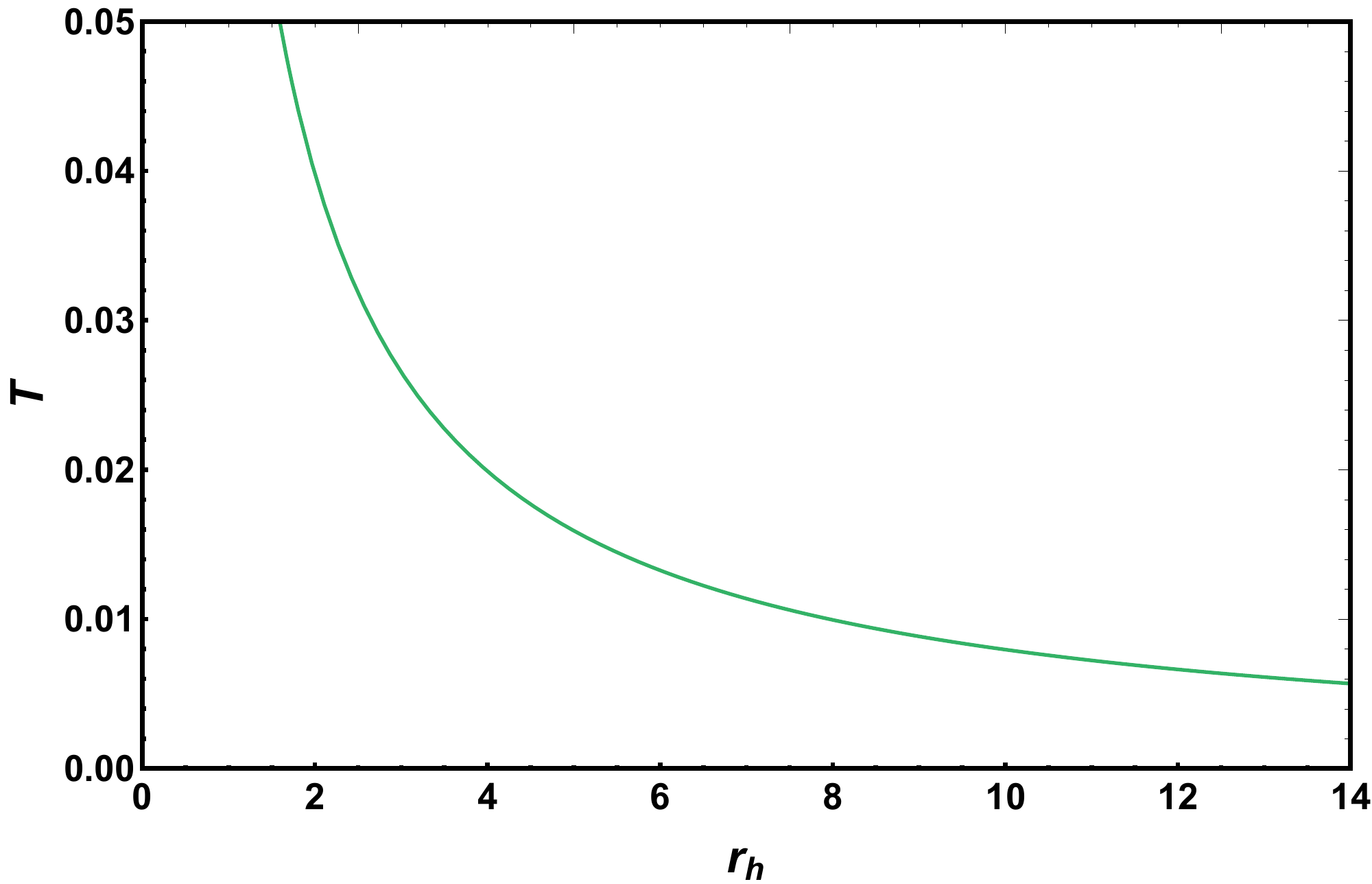} \hspace{1cm}
\caption{\label{4} The temperature function $T(r_h)$ of the Schwarzschild black hole shown in the $T-r_h$ plane.}
\end{figure}

Schwarzschild is a static spherically symmetric black hole without charges and cosmological constant. Its metric function is
\begin{align}
f = 1 - \frac{2M}{r} ,
\end{align}
and the temperature
\begin{align}
T = \frac{1}{4\pi r_h} .
\end{align}
As shown in Fig.~\ref{4}, the temperature function of the Schwarzschild black hole is monotonically decreasing; therefore, the Schwarzschild black hole corresponds to the $B^-$ class.

\subsubsection{Kerr--AdS}
The Kerr--AdS black hole is a four dimensional rotating black hole with a negative cosmological constant. Its metric is given by
\begin{widetext}
\begin{align}
ds^2 = -\frac{\Delta}{\rho^2} \left( dt - \frac{a \sin^2 \theta}{\Xi} d\varphi \right)^2 + \frac{\rho^2}{\Delta} dr^2 + \frac{\rho^2}{\Sigma} d\theta^2 + \frac{\Sigma \sin^2 \theta}{\rho^2} \left[ a dt - \frac{(r^2 + a^2)}{\Xi} d\varphi \right]^2,
\end{align}
\end{widetext}
where
\begin{align}
\Delta = \left( r^2 + a^2 \right) \left( 1 + \frac{r^2}{\ell^2} \right) - 2mr, \quad \Xi = 1 - \frac{a^2}{\ell^2},
\end{align}
and
\begin{align}
\rho^2 = r^2 + a^2 \cos^2 \theta, \quad \Sigma = 1 - \frac{a^2}{\ell^2} \cos^2 \theta.
\end{align}
Where $a$ is the rotational parameter. The expression for the angular momentum is
\begin{align}
J = \frac{m}{\Xi^2} a.
\end{align}
We introduce the following dimensionless rescaling (which does not change the essential behavior of the temperature function $T(r_h)$)
\begin{align}
\tilde{r}_h = \frac{r_h}{\ell},~\quad \tilde{a} = \frac{a}{\ell},~\quad \tilde{m} = \frac{m}{\ell},~\quad \tilde{J} = \frac{J}{\ell^2},~\quad \tilde{T} = T\ell.
\end{align}
At the critical point, $\tilde{J}_c=0.0239$ and $\tilde{r}_{hc}=0.4588$.

\begin{figure}
\centering
\includegraphics[width=0.4485\textwidth, height=0.22425\textheight]{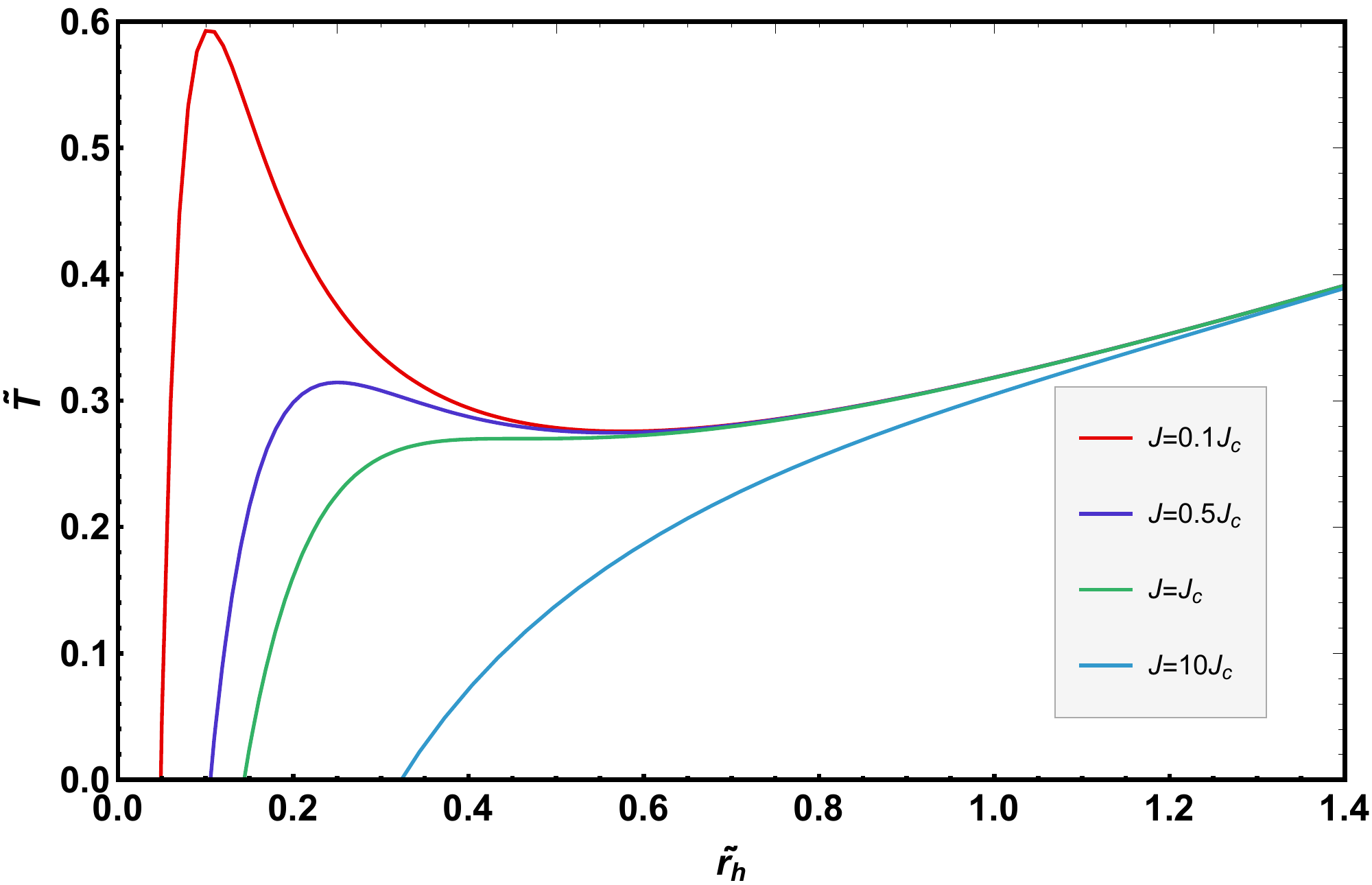} \hspace{1cm}
\caption{\label{5} The temperature function $\tilde{T}(\tilde{r}_h)$ of the Kerr--AdS black hole shown in the $\tilde{T}-\tilde{r}_h$ plane for different values of $\tilde{J}$.}
\end{figure}

For the Kerr--AdS black hole, we work in the canonical ensemble with fixed angular momentum $\tilde{J}$. As shown in Fig.~\ref{5}, as $\tilde{J}$ decreases, the Kerr--AdS black hole transitions from the $B^+$ class to the $A2$ class.

Note that when $\tilde{J} < \tilde{J}_c$, the temperature function $\tilde{T}(\tilde{r}_h)$ exhibits a double extremum structure, whereas when $\tilde{J} > \tilde{J}_c$, $\tilde{T}(\tilde{r}_h)$ is monotonically increasing. This implies that the Kerr--AdS black hole cannot belong to the $A1^+$ class or the $B^-$ class. A similar situation holds for the Schwarzschild--AdS black hole.

Furthermore, because the temperature $\tilde{T}$ of a rotating black hole is implicitly related to the angular momentum $\tilde{J}$  and the horizon radius $r_h$, it is difficult to obtain an analytical discriminant for the equation $\tilde{T}' = 0$. Nevertheless, by numerically plotting the $\tilde{T}(\tilde{r}_h)$ curve and counting the number of extremal points, an accurate classification can still be achieved, which demonstrates the flexibility of the local framework.

\subsection{Summary table}
We summarize the various black holes investigated in this paper and their corresponding classifications in Table~\ref{st}.

\begin{table*}[htbp]
\centering
\setlength{\tabcolsep}{3.2pt}
\begin{tabular*}{\textwidth}{c@{\extracolsep{\fill}}cccccc}
\hline\hline
Local class & 1st order & $W$ & Global class & Order & Riemann surface & Black hole system\\
\midrule
$A2$     & Yes & $+1$ & $W^{1+}$ & $[+, -, +]$ & 3 foliations  & \makecell{Hayward--AdS (small $\tilde{g}$)\\RN--AdS (small $Q/\ell$)\\Kerr--AdS (small $\tilde{J}$)}\\
\midrule
$A1^{+}$  & No  & $0$  & $W^{0+}$ & $[+, -]$    & 2 foliations & Hayward\\
\midrule
$A1^{-}$  & No  & $0$  & $W^{0-}$ & $[-, +]$    & 2 foliations & Schwarzschild--AdS\\
\midrule
$B^{+}$   & No  & $+1$ & $W^{1+}$ & $[+]$       & 1 foliation  & \makecell{Hayward--AdS (large $\tilde{g}$)\\RN--AdS (large $Q/\ell$)\\Kerr--AdS (large $\tilde{J}$)\\Schwarzschild--AdS (small $\ell^2$)}\\
\midrule
$B^{-}$   & No  & $-1$ & $W^{1-}$ & $[-]$       & 1 foliation  & Schwarzschild\\
\bottomrule
\end{tabular*}
\caption{Summary of black hole systems and their classifications. Here, ``1st order'' denotes the first-order phase transition. For each black hole system considered in this paper, we list its local geometric classification, whether it exhibits a first-order phase transition, its topological number $W$, its global topological classification, its winding number order, the number of foliations of the corresponding Riemann surface, and representative examples of parameter choices or black hole models that fall into each class.}
\label{st}
\end{table*}

\section{Further comments and conclusions}

In this paper, we establish a unified framework that connects three representative classification schemes in black hole thermodynamics: the classification based on locally defined geometric properties, the classification based on global topological invariants, and the classification based on the number of Riemann surface foliations in complex plane. On the basis of the two dictionaries constructed, we prove that the number of extremal points $n$ on the black hole temperature curve simultaneously determines the classification in the local scheme, the winding number order and the total topological number in the global scheme, and the number of Riemann surface foliations in the complex analysis framework. The mathematical origin of this unified picture lies in the critical point structure of the black hole solution space; it is precisely the emergence of these fold singularities that leads to the multivaluedness of physical or geometric quantities, the generation of topological defects, and the multi-foliations structure of the Riemann surface associated with the complex analytic function. This framework provides a unified basis for classifying black hole systems in quantum gravity and other gravitational theories.

The temperature function encodes not only thermodynamic properties but also the underlying geometric structure of the solution space. Consequently, for any black hole system, the thermodynamic analysis should begin with investigating its temperature function. As a direct application of this unified framework, we take the RN--AdS black hole as an example to demonstrate its simplicity and predictive power: by simply plotting the temperature curve and counting the number of extremal points, one can immediately read off the topological number, winding number order, number of Riemann surface foliations, and whether a first-order phase transition occurs. This greatly streamlines the analysis of black hole thermodynamics and provides a unified theoretical tool for subsequent investigations of more complex black hole systems, such as rotating black holes, higher dimensional black holes, and black holes in modified gravitational theories.

It is noteworthy that Ref.~\cite{Wei:2024gfz} pointed out that when the asymptotic behavior of a black hole does not satisfy the standard conditions (such as in multi-charged AdS black holes), a topological phase transition may occur, meaning that the total topological number of the system varies with parameters. This provides a natural direction for extending the unified framework established in this paper: does the correspondence among the number of extremal points on the temperature curve, the winding number order, and the number of Riemann surface foliations still hold in more general black hole systems? If not, how do these changes reconcile with one another? Furthermore, investigating systems with a larger number of extremal points (such as $6$-dimensional Gauss-Bonnet black holes) will reveal richer phase transition structures. Exploring these questions will further advance our understanding of the unified thermodynamic, geometric, and topological picture of black holes.

In summary, the unified perspective established in this paper reveals the universal geometric origin of black hole thermodynamic properties: whether manifested as multivaluedness, topological numbers, or the foliation structure of Riemann surfaces, these diverse behaviors ultimately trace back to the fold singularities within the solution space. Different black hole solutions, governed by distinct Einstein equations derived from different actions, naturally exhibit or lack such fold singularities, which in turn determine their thermodynamic characteristics. This understanding advances the study of black hole thermodynamics from mere classification to a deeper comprehension of the roots of classification, and provides a theoretical foundation for further exploring analogous structures in quantum gravity.

\section*{Appendix: $\tilde{T}(\tilde{r}_h)$ curve of the 6-dimensional charged Gauss-Bonnet--AdS black hole}
In this appendix, we discuss the $\tilde{T}(\tilde{r}_h)$ curve of the 6-dimensional charged Gauss-Bonnet--AdS (6D CGB--AdS) black hole as a typical example with more number of extrema.

\begin{figure*}
	\begin{minipage}{1\hsize}
		\begin{center}
			
			\subfigure[]{
				\includegraphics*[scale=0.195]{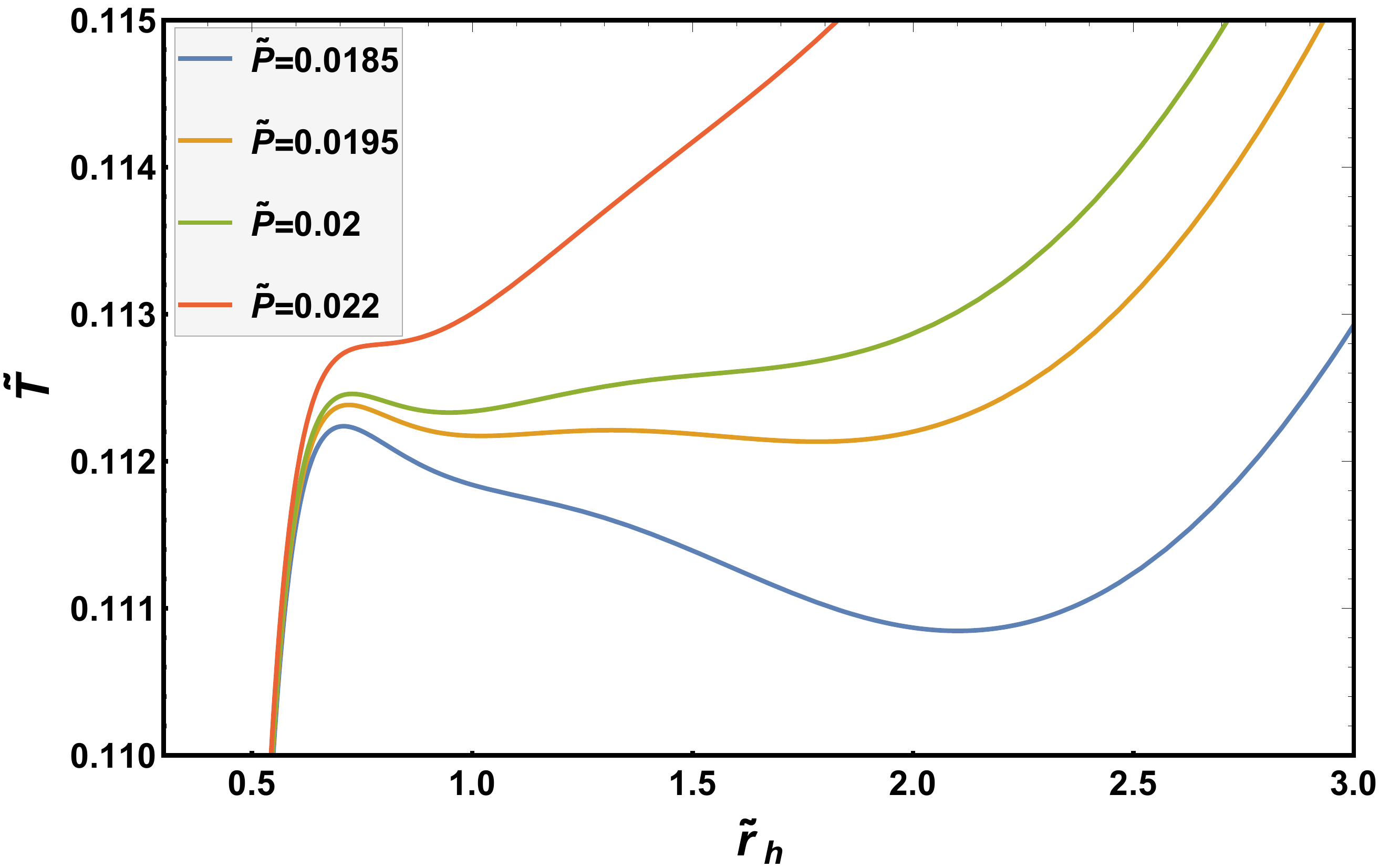}}
			\subfigure[]{
				\includegraphics*[scale=0.195]{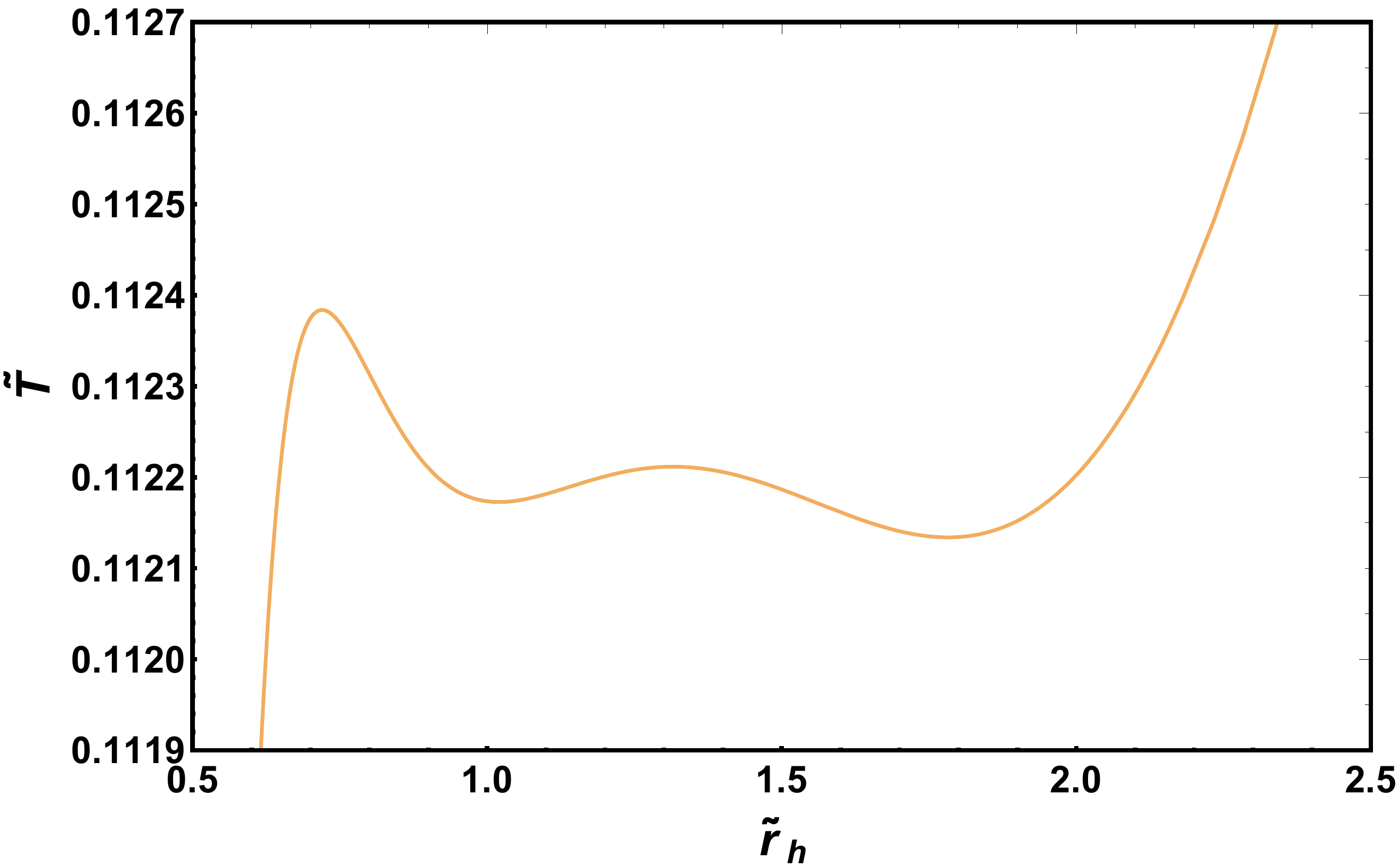}}
            
		\end{center}
		\caption{The temperature function $\tilde{T}(\tilde{r}_h)$ of the 6D CGB--AdS black hole shown in the $\tilde{T}-\tilde{r}_h$ plane for different values of $\tilde{P}$. (a) $\tilde{T}(\tilde{r}_h)$ curves for four different values of $\tilde{P}$, (b) $\tilde{T}(\tilde{r}_h)$ curve at $\tilde{P}=0.0195$.
        }
		\label{6}
	\end{minipage}
\end{figure*}

The temperature of the 6D CGB--AdS black hole is given by \cite{Wei:2014hba}
\begin{align}
T = \frac{8\pi P r_h^8 + 6r_h^6 + 2\alpha r_h^4 - Q^2}{8\pi r_h^7 + 16\pi  \alpha r_h^5},
\end{align}
where the pressure $P=\frac{5}{4\pi \ell^2}$ (it
has dimensions of $[\text{length}]^{-2}$), $r_h$ is the horizon radius (it
has dimensions of $[\text{length}]$), $\alpha$ is the coupling constant (it
has dimensions of $[\text{length}]^{2}$), and $Q$ is the charge (it
has dimensions of $[\text{length}]^{3}$). We introduce the dimensionless rescaling
\begin{align}
\tilde{r}_h = \frac{r_h}{\sqrt{\alpha}},~\tilde{T} = T\sqrt{\alpha},~\quad \tilde{P} = P\alpha,~\quad \tilde{Q} = \frac{Q}{\alpha^{\frac{3}{2}}}.
\end{align}
In the subsequent analysis, for consistency with Ref.~\cite{Wei:2014hba}, we set $\alpha = 1$. In the canonical ensemble, the $\tilde{T}(\tilde{r}_h)$ curve of the 6D CGB--AdS black hole is shown in Fig.~\ref{6}. We select four values of the pressure $(0.0185,\,0.0195,\,0.02,\,0.022)$ for the analysis. As a black hole exhibiting richer phase structure under specific parameters \cite{Wei:2014hba}, the 6D CGB--AdS black hole, in the canonical ensemble with $\tilde{Q}=0.18$, possesses four extrema (five branches) in the $\tilde{T}(\tilde{r}_h)$ curve at $\tilde{P}=0.0195$, consistent with the result from the complex analysis framework \cite{Xu:2023vyj}. Meanwhile, according to the dictionary established in this work, the 6D CGB--AdS black hole ($\tilde{P}=0.0195$) corresponds, in the topological framework, to the winding number order $[+,\,-,\,+,\,-,\,+]$ and the topological number $W=+1$.

It is worth noting that the local framework also allows us to obtain the conclusion of four extrema algebraically. From $T'=0$ we can determine the number of real roots, which yields
\begin{widetext}
\begin{align}
\frac{8r_h^3 + 36r_h^5 + 64P\pi r_h^7}{16\pi r_h^5 + 8\pi r_h^7} - \frac{(80\pi r_h^4 + 56\pi r_h^6)(-Q^2 + 2r_h^4 + 6r_h^6 + 8P\pi r_h^8)}{(16\pi r_h^5 + 8\pi r_h^7)^2} = 0.\label{DT}
\end{align}
\end{widetext}
For $\tilde{P}=0.0195$ and $\tilde{Q}=0.18$, Eq.~(\ref{DT}) admits four positive real roots:
\begin{align}
r_{h1} = 0.7199,\,r_{h2} = 1.0200,\,r_{h3} = 1.3155,\,r_{h4} = 1.7823.
\end{align}
This enables a more rapid parameter scan of more complex black holes, thereby facilitating their thermodynamic classification.

\acknowledgments
This work was supported by the National Natural Science Foundation of China (Grants Nos. 12533001, 12575049, and 12473001), the National SKA Program of China (Grants Nos. 2022SKA0110200 and 2022SKA0110203), the China Manned Space Program (Grant No. CMS-CSST-2025-A02), and the 111 Project (Grant No. B16009).

\bibliography{bh_unify}

\end{document}